# New possibilities in the utilization of holographic screens

José J. Lunazzi

Universidade Estadual de Campinas, Instituto de Física

C.P. 6165, 13081 Campinas-SP- Brazil

## ABSTRACT

Directionality in a holographic screen may be useful for projecting images to be seen in complete horizontal parallax. The continuous sequence of views from an object may be transferred from the object and enlarged at the screen giving the same appearance of a holographic image. Due to the actual movement of the object, images in four dimensions may now be produced with a projector whose spatial dimensions are much smaller than those of the screen. One technique that allows for this result is the chromatic encoding of views by means of holographic optical elements. This is demonstrated to be a complete reproduction of the original light distribution, although its spectral continuous distribution of colors makes each view monochromatic, not allowing for a good color reproduction. The resulting system may substitute conventional holography in some visual applications were registering is not necessary but only the effect of the phantasmagoric image. Furthermore, it allows for the enlarging of holographic images, performing also the direct conversion of the image of a conventional off-axis hologram to a white-light image.

## 1. INTRODUCTION

Holographic screens had been proposed for the purpose of avoiding the use of stereo glasses in a stereo projection [1,2] but its possibilities for displaying the continuous parallax were not described up to date. Since displaying only the horizontal parallax may suffice for a good perception of the 3D realm of the scene, as in a Benton hologram [3], this is the only case that we consider in this article. One of the possibilities was reported [4], and is derived from the chromatic encoding of views that takes place in a natural way when a hologram or a diffraction grating are used for imaging under white-light illumination [5,6]. By considering that the diffraction process allows for the imaging of each view from the scene in a single wavelength, while distributing the continuous sequence of views in the same sequence than the spectral wavelengths, we developed a holographic screen that matches this phenomena allowing for the perfect decoding of this synthetic image. The image originated from each view from the object is made to propagate in a direction that corresponds to its original direction, creating around the screen a light distribution that resembles very well the original one that exists around the object. As a first approach for describing the situation, we must consider that a diffracting grating of the same period than the one employed for encoding is capable of decoding the image from the original scene. A symmetric situation may be created by means of the object, the two gratings and the projecting lens allowing for the generation of a pseudoscopic real image and an orthoscopic virtual image [7]. But this case does not allow for obtaining an image that may be larger than the projecting system itself. It is necessary for this to give to the screen the focusing properties of a holographic lens, as it will be described in the following sections of this paper. Just for simplicity, we consider the situation as corresponding to two views, two wavelengths, prior to generalizing to the continuous case of spectral distribution.

## 2. THE CHROMATIC ENCODING OF VIEWS

The close relationship that exists between the stereographic registering of a scene and the photographic registering of the same scene after diffraction through a grating (see Ref.5, eqs.7,8) indicates that if we project the image points on the screen making them visible only to the corresponding right or left eye, the appearance of the scene will be that of a stereo scene. We may perform this by means of a directional screen, as follows.

## 3. THE HOLOGRAPHIC LENS EMPLOYED AS A DIRECTIONAL SCREEN

A convergent holographic lens is made by means of two laser beams of wavelength $\lambda$, one diverging from point o, the other converging to the point i. By also indicating as o and i the distances from this points to the holographic plate, an imaging relationship exists in terms of the focal length f:

$$f^{-1} = o^{-1} + i^{-1} \tag{1}$$

Which may be generalized for the case of a different illuminating wavelength by the expression:

$$f^{-1} \lambda_R^{-1} \lambda_{RC} = s^{-1} + s'^{-1} \tag{2}$$

s, s' being the object and image distances respectively. Not only the focusing distance changes according to wavelenght but also the lateral displacement of the image, according to the fundamental equation of a diffraction grating, giving a light distribution as represented in Fig 1, for two different wavelengths.

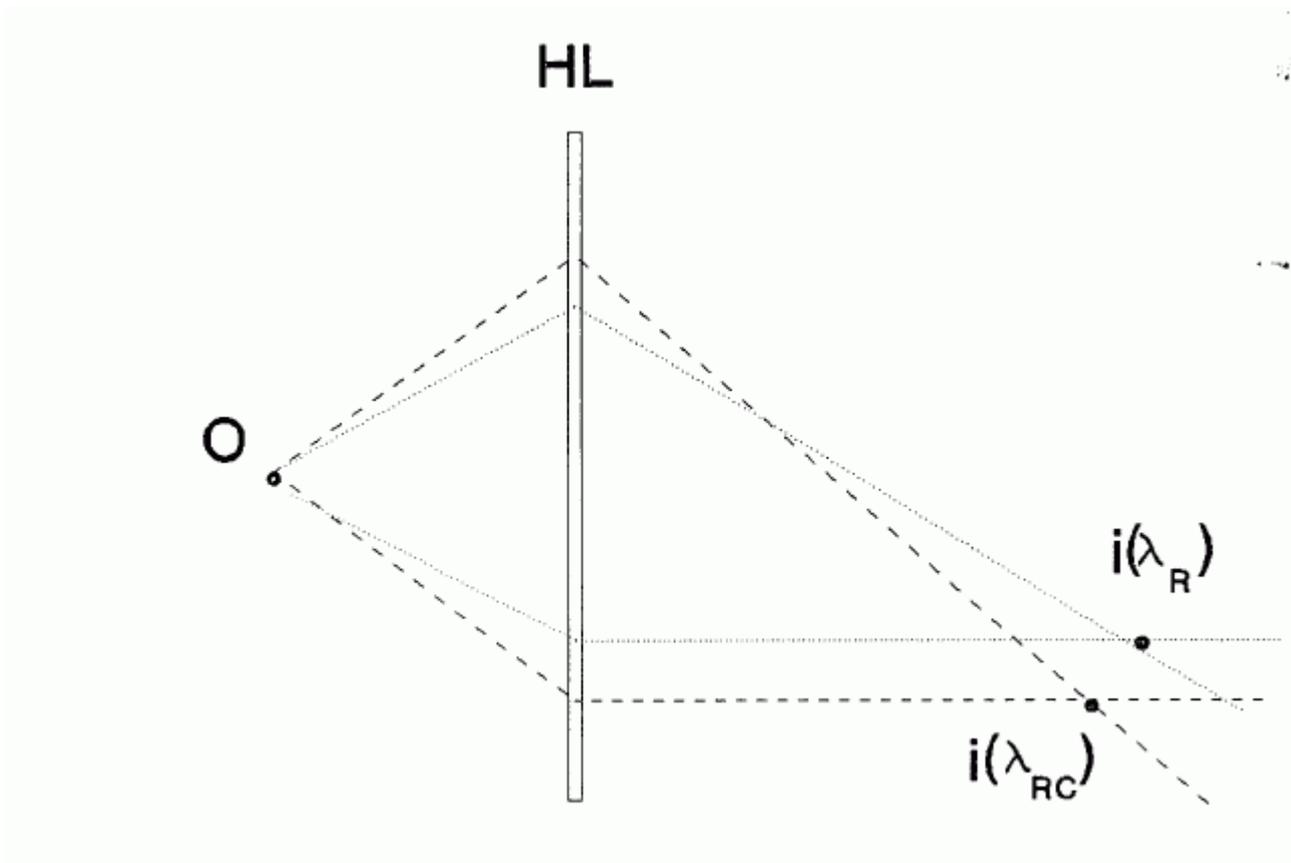

Fig 1: Light distribution for two different wavelengths

They were indicated by dashed lines, the length of its segments being related to its corresponding wavelengths. This figure may also be considered as an useful scheme for the continuous case of imaging under white light. When a multi-wavelength image is projected onto the holographic optical element **HL** by means of a lens **L**, the result is that we may see each wavelength component of the figure from a different point of view.

## 4. THE HOLOPROJECTION CASE

For the case of chromatic encoding by diffraction, it lends to the situation of Fig 2, **o** being a white object illuminated under white light and **DG** a holographic diffraction grating. The observer may then see each view from the object in a representation that matches the original seeing of it, a situation that may be denominated as "holoimage", having the realm of a holographic image.

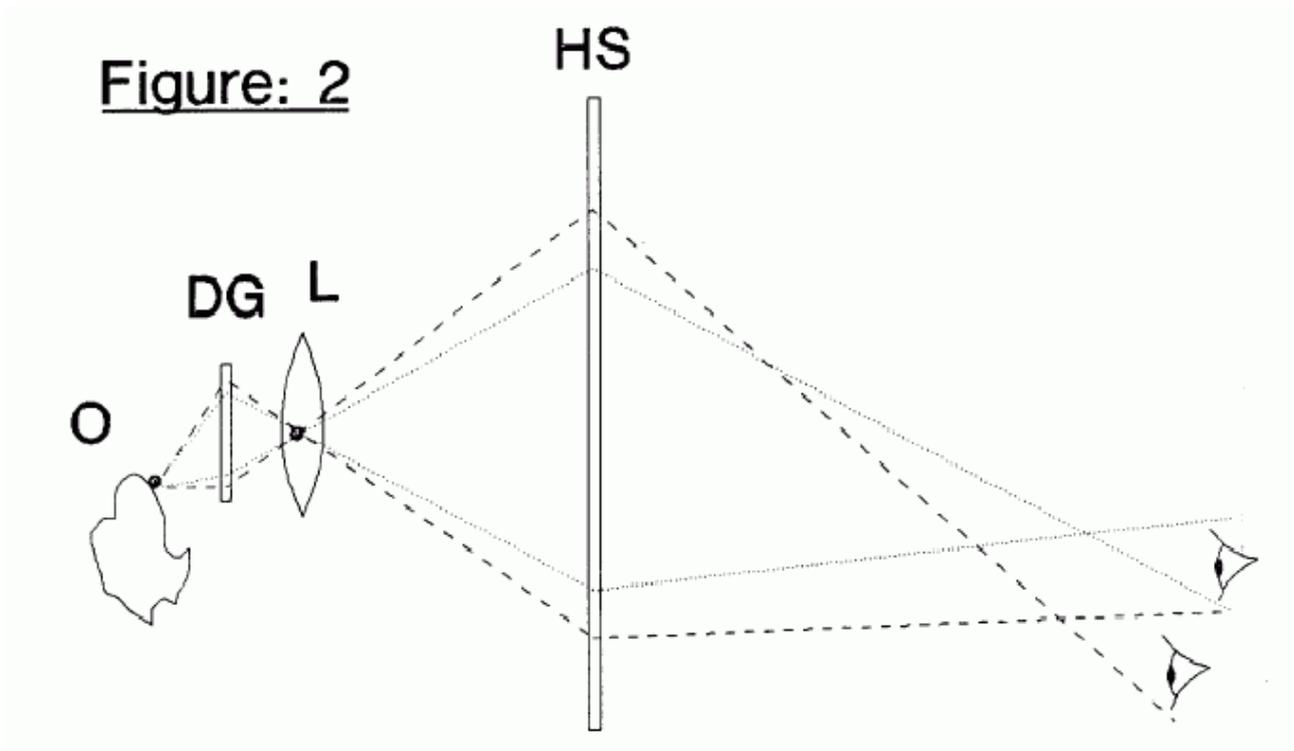

Fig 2: Chromatic encoding by diffraction

## 5. THE ENLARGING AND WHITE LIGHT CONVERSION OF A HOLOGRAM

As it was described in (5), the illumination in white light of a simple of-axis hologram whose reference bean impinges from a lateral position allows for the chromatic encoding of views. It suffices then to reconstruct such a hologram by means of a small filament lamp for obtaining its holographic visualization when the image is projected on a holographic screen. The common chromatic blurring of the image is the converted into continuous horizontal parallax for the image at the screen.

## 6. THE NEED FOR A PROJECTING LENS

In our description, the projecting lens was, for the sake of simplicity, considered as a pinhole lens, an element of much reduced dimensions whose purpose is to define a possible path for the light rays. The need for a lens is the same than in the common imaging case| more light is needed for increasing the brightness at the image while reducing diffraction aberrations. We obtain both at the price of a reduced depth of field. The geometrical aberrations of this lens appear as the most important for defining the quality of the image.

## 7. CONSIDERATIONS ON THE ACHIEVABLE DIFFRACTION EFFICIENCY OF THE HOE`s

The problem of luminous efficiency of the double diffraction system depends on the diffraction efficiency of the coding and decoding elements. Due to the need of using the whole visible spectrum, high efficiencies are limited by the wavelength selective effects of Bragg diffraction.

## 8. EXPERIMENTAL DETAILS

Many situations were experimented for projecting the images from small objects. Different periods for the encoding grating and for the holographic screen were tried. The most remarkable result was obtained by illuminating an object whose width, depth and height dimensions were 2 cm, 5 cm and 4 cm respectively, using a halogenous 250 W projector lamp, a 1000 lines/mm transmission holographic grating, a f= 350 mm, f| 5,6 projecting objective and a holographic screen 30 cm large by 40 cm high constructed by two interfering beams that impinged the holographic emulsion at a relative angle of $45^0$. Although the grating periods used for encoding and decoding were not precisely the same, the result was that we could see a clear image floating at 30 cm in front of the screen in perfect continuous horizontal parallax, looking exactly as a holographic image. A lateral field of view of 33 cm was obtained at 90 cm distance from the screen, which was at a 180 cm distance from the projecting lens. For enlarging and converting a hologram we made a simple off-axis hologram in 35mm SO-253 Kodak holographic film, illuminating it with the direct filament of a halogeneous lamp whose dimensions were smaller than 1 mm in diameter, at a distance of 1 m. The holographic image was projected on a holographic screen 15 cm wide and 30 cm high, which corresponds to an enlarging factor of about 10. The image had a depth of field of about 50 cm, being very clear in depth and parallax although not very bright and with some speckle noise. In both cases, the images were less bright than conventional holographic images, requiring to be observed in a dark ambience.

## 9. CONCLUSIONS

A new imaging system was developed that allows for keeping the main three-dimensional characteristics of a white-light luminous field, allowing for enlarging its dimensions by means of conventional projecting objectives. It may be useful for substituting conventional holograms in visual arts in the cases where the registering of the scene is not necessary. The addition of movement to the image increases very much its visual effect, offering to the public the impression of the holographic movies of the future. The capability of enlarging holograms may also be an interesting application, since it allows for using a small embossed hologram as a large size one, giving an economy of material comparable to the area enlarging factor of the imaging system.

## 10. ACKNOWLEDGEMENTS

This work has been made possible by support from the Found of Assistance to Research and Teaching-FAEP of the Campinas State University-UNICAMP, the Foundation of Assistance to Research of the Sao Paulo State-FAPESP and the National Council of Research-CNPq.
Thanks to Daniel S. F. Magalhães for help in formatting this article to ArXiv.